\documentclass[pdflatex,sn-nature]{sn-jnl}

\usepackage{graphicx}%
\usepackage{multirow}%
\usepackage{amsmath,amssymb,amsfonts}%
\usepackage{amsthm}%
\usepackage{mathrsfs}%
\usepackage[title]{appendix}%
\usepackage{xcolor}%
\usepackage{textcomp}%
\usepackage{manyfoot}%
\usepackage{booktabs}%
\usepackage{algorithm}%
\usepackage{algorithmicx}%
\usepackage{algpseudocode}%
\usepackage{listings}%
\usepackage{subcaption}
\usepackage{url}%
\theoremstyle{thmstyleone}%
\theoremstyle{thmstyletwo}%

\theoremstyle{thmstylethree}%

\begin{document}

\title[SEPNET-PRISM-R]{Real-Time 24-hour Ahead Forecasting of Solar Energetic Particle Events with \texttt{SEPNET-PRISM-R}}

\author[1]{\fnm{Yian} \sur{Yu}}
\author*[1]{\fnm{Yang} \sur{Chen}}\email{ychenang@umich.edu}
\author[2]{\fnm{Lulu} \sur{Zhao}}
\author[3]{\fnm{Kathryn} \sur{Whitman}}
\author[2]{\fnm{Ward} \sur{Manchester}}
\author[2]{\fnm{Tamas} \sur{Gombosi}}
\affil*[1]{\orgdiv{Department of Statistics}, \orgname{University of Michigan}, \orgaddress{\city{Ann Arbor}, \state{MI}, \country{USA}}}
\affil[2]{\orgdiv{Department of Climate and Space Sciences and Engineering}, \orgname{University of Michigan}, \orgaddress{\city{Ann Arbor}, \state{MI}, \country{USA}}}
\affil[3]{\orgdiv{NASA Space Radiation Analysis Group}, \orgname{Johnson Space Center}, \orgaddress{\city{Houston}, \state{TX}, \country{USA}}}


\abstract{Solar energetic particle (SEP) events, operationally defined as proton flux exceeding 10 pfu in the $>$10 MeV channel, threaten spacecraft systems, astronaut safety, and high-latitude aviation, yet continuous real-time forecasting with adequate lead time remains difficult. We present \texttt{SEPNET-PRISM-R}, a near-real-time system that issues hourly, 24-hour-ahead probabilistic SEP forecasts. The model integrates multimodal solar data—including active-region patches, flare features, coronal mass ejection (CME) characteristics, and historical proton flux—to concurrently predict SEP occurrences, peak proton flux, soft X-ray flux, and CME velocity, with the primary objective being SEP forecasting. Building on existing literature, \texttt{SEPNET-PRISM-R} enhances operational performance by incorporating a post-processing module that enables cross-validated decision-making and interpretable quantification of feature importance. The training dataset spans from February 1986 to September 2025, with subsequent forecasts generated sequentially through the present via a real-time space weather prediction platform that provides predictive probabilities, feature attributions, and historical analogs, enabling more reliable and informed operational decision-making. The proposed framework substantially suppresses false-alarm rates while maintaining robust event-detection capabilities, as evidenced by case studies and aggregate statistics.  
}


\maketitle

\section{Introduction}\label{sec1}
Solar energetic particle (SEP) events are among the most hazardous manifestations of solar activity and represent a major component of space weather \citep{reames1999particle}. During solar eruptions, protons and heavier ions can be accelerated to relativistic energies through magnetic reconnection associated with solar flares and through shock waves driven by fast coronal mass ejections (CMEs) \citep{desai2016large,Zhao2024SOFIE}. These particles can propagate through the heliosphere and reach Earth within tens of minutes to several hours after eruption, and the flux may remain elevated for days. Operationally, NOAA (National Oceanic and Atmospheric Administration) defines an SEP event when the integral proton flux in the GOES $>$ 10 MeV channel exceeds 10 particle flux units (pfu), a threshold now standard in space weather monitoring and forecasting. SEP events threaten spacecraft electronics, satellite operations, astronaut radiation safety, high-latitude aviation, and radio communications \citep{schwadron2010earth,shea2012space}, making reliable forecasting an important objective for both scientific research and operational space weather services \citep{Whitman20235161,kasapis2026reviewMLSEPmodels}. 

However, SEP forecasting remains scientifically difficult, as event occurrence depends on a combination of solar eruption characteristics, the underlying particle acceleration mechanism, interplanetary transport, and the magnetic connectivity between the acceleration site and the observer \citep{desai2016large}. Large X-class flares often fail to produce measurable SEP enhancements at Earth when the associated magnetic geometry is unfavorable, while moderate eruptions can occasionally generate intense particle events under efficient magnetic connection \citep{Kahler1982roleFlare}. Fast CMEs are similarly important for large SEP events, yet CME speed alone is insufficient for reliable prediction, since shock formation and subsequent particle transport also depend on the ambient solar wind and the evolving heliospheric magnetic field \citep{gopalswamy2002interacting}. 

A wide range of forecasting approaches has been proposed over the past several decades. Early empirical methods primarily relied on flare magnitude and location, and CME speed to estimate SEP probability \citep{Laurenza2009}. Although physically interpretable, these methods are constrained by simplifying assumptions and struggle to capture complex nonlinear interactions among predictors. Machine learning techniques have become increasingly popular for SEP forecasting \citep{camporeale2018machine}, various statistical and deep learning models incorporating magnetic field measurements, flare characteristics, CME properties, radio burst observations, and historical particle fluxes to improve skill \citep{papaioannou2018nowcasting,Kasapis2022SW}. These data-driven approaches have delivered measurable gains in forecast performance, underscoring their promise for operational space weather forecasting.

Several obstacles nevertheless stand between these advances and a successful operational deployment \citep{Camporeale2019}. Many published studies rely on retrospectively constructed datasets in which training and testing samples are randomly partitioned~\citep{Kasapis2022SW,yu2026SEPNET-PRISM}. Because neighboring observations are often highly correlated in time, such an evaluation protocol can substantially inflate the predictive skill scores. Furthermore, many studies focus primarily on event classification, while paying less attention to operational considerations, such as continuous hourly updates, forecast stability, false-alarm control, computational cost, and model interpretability~\citep{roberts2017cross}. Closing the gap between retrospective evaluation and real-time operation is especially important for SEP prediction, considering its potential impacts on near-Earth environments. 

The previously developed \texttt{SEPNET} family of deep-learning forecasting models partially addressed these issues. The initial operational version, \texttt{SEPNET-O} \citep{yu2026SEPNET}, integrated SHARP magnetic parameters with flare characteristics to produce probabilistic forecasts of SEP events over the next 24 hours. The subsequent methodological refinement, \texttt{SEPNET-PRISM} \citep{yu2026SEPNET-PRISM}, is a multitask extension trained on a substantially larger historical dataset spanning more than two solar cycles with additional auxiliary targets, including future proton flux, GOES soft X-ray flux, and CME speed. It was shown that the joint learning of these related quantities improved predictive performance and enriched the learned physical representation beyond binary classification. However, \texttt{SEPNET-PRISM} centered on model development and retrospective evaluation rather than real-time operation. While \texttt{SEPNET-O} was shown real-time capability and has been running live on NASA CCMC (Community Coordinated Modeling Center)~\citep{nasa_ccmc_sepnet}, its outputs remained sensitive to short-term fluctuations from noisy observations, producing unstable warning behavior under sequential prediction. For instance, during the Artemis II space mission, the live forecasting produced by \texttt{SEPNET-O} issued false positive warnings of transient non-SEP soarl eruptions with a significantly perturbed prediction probability, which are primarily caused by noisy observations and random fluctuations. Furthermore, neither \texttt{SEPNET-O} nor \texttt{SEPNET-PRISM} had been paired with the temporal post-processing, independent post-training evaluation, or operational interpretability that a desirable operational system would have.

Recognizing the substantial gaps in methodology and the lack of practical real-time decision-supporting for operational forecasting, we propose \texttt{SEPNET-PRISM-R}, a thorough extension of \texttt{SEPNET-PRISM} built specifically for continuous near-real-time forecasting. With a trained model, forecasts are issued sequentially every hour using only the observations available at issuance time, closely mirroring true operational deployment. To improve temporal consistency, \texttt{SEPNET-PRISM-R} adopts the Hidden Semi-Markov Model (HSMM) \citep{yu2010hidden} as a post-processing layer that explicitly models the persistence of SEP and quiet states, reducing isolated false alarms driven by short-term probability fluctuations. We evaluate the resulting system over a post-training period from September 2025 through July 2026 (that has no overlap with training data) to obtain a realistic picture of operational performance. We use SHapley Additive exPlanations (SHAP) \citep{Daniel2022GBM,wang2024feature} to quantify global and forecast-specific feature contributions, improving transparency and offering physical insight into the drivers of each operational prediction. In our interactive online platform, we show real-time forecasting results, together with intuitive explanations of feature contributions and matched historical event visualizations, to facilitate interpretation of prediction results in terms of both reliability, stability and explanability. 

\section{Results}\label{sec2}

This section evaluates the forecasting performance of \texttt{SEPNET-PRISM-R} for hourly 24-hour-ahead operational SEP forecasting and compares it with the previous operational model \texttt{SEPNET-O}. We begin by evaluating the model's overall performance across the full test interval. Subsequently, we examine detailed case studies of SEP and non-SEP events, present the auxiliary regression results, and analyze feature importance using SHAP values. Finally, we demonstrate the near-real-time hourly web dashboard that delivers these operational outputs alongside the analog-based context established in Section \ref{sec:analog}. 
Classification performance is evaluated using accuracy (ACC), the area under the receiver operating characteristic curve (AUC) \cite{muschelli2020roc}, the false positive rate (FPR), the F1 score \cite{lipton2014f1}, the probability of detection (POD) \cite{wehling2011probability}, the false alarm ratio (FAR) \cite{macmillan1985detection}, the true skill statistic (TSS) \cite{doswell1990summary}, and the Heidke skill score (HSS), in accordance with established practices in space weather forecast verification \cite{Leka2019}. Regression performance is quantified using the root mean square error (RMSE), the mean absolute error (MAE), and the coefficient of determination (R$^2$). 

\subsection{Overall Forecast Performance}\label{subsec2-1}
\begin{figure*}[htbp]
    \centering

    \begin{subfigure}[t]{1\textwidth}
        \centering
        \includegraphics[width=\textwidth]{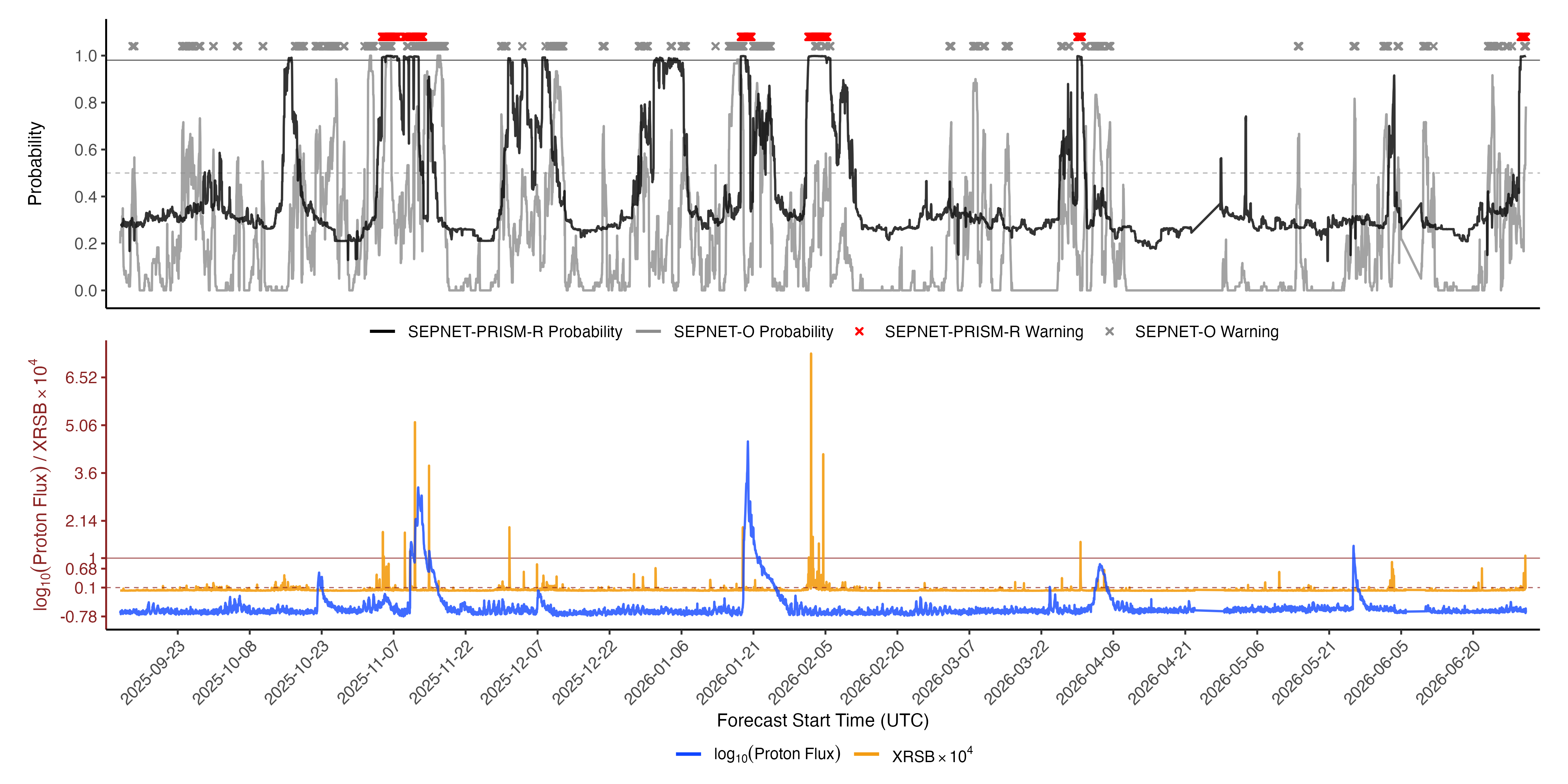}
        \caption{Operational forecasting results from September 10, 2025, to July 1, 2026. The upper portion shows forecast probabilities from \texttt{SEPNET-PRISM-R} (black) and \texttt{SEPNET-O} (gray), together with their corresponding warning indicators (red and gray crosses on the top). The lower portion shows the observed solar and particle context, including $\log_{10}$ proton flux (blue) and XRSB soft X-ray flux scaled by $10^4$ (orange).}
        \label{fig:ProbsPFXRSB}
    \end{subfigure}
    \begin{subfigure}[t]{1\textwidth}
        \centering
        \includegraphics[width=\textwidth]{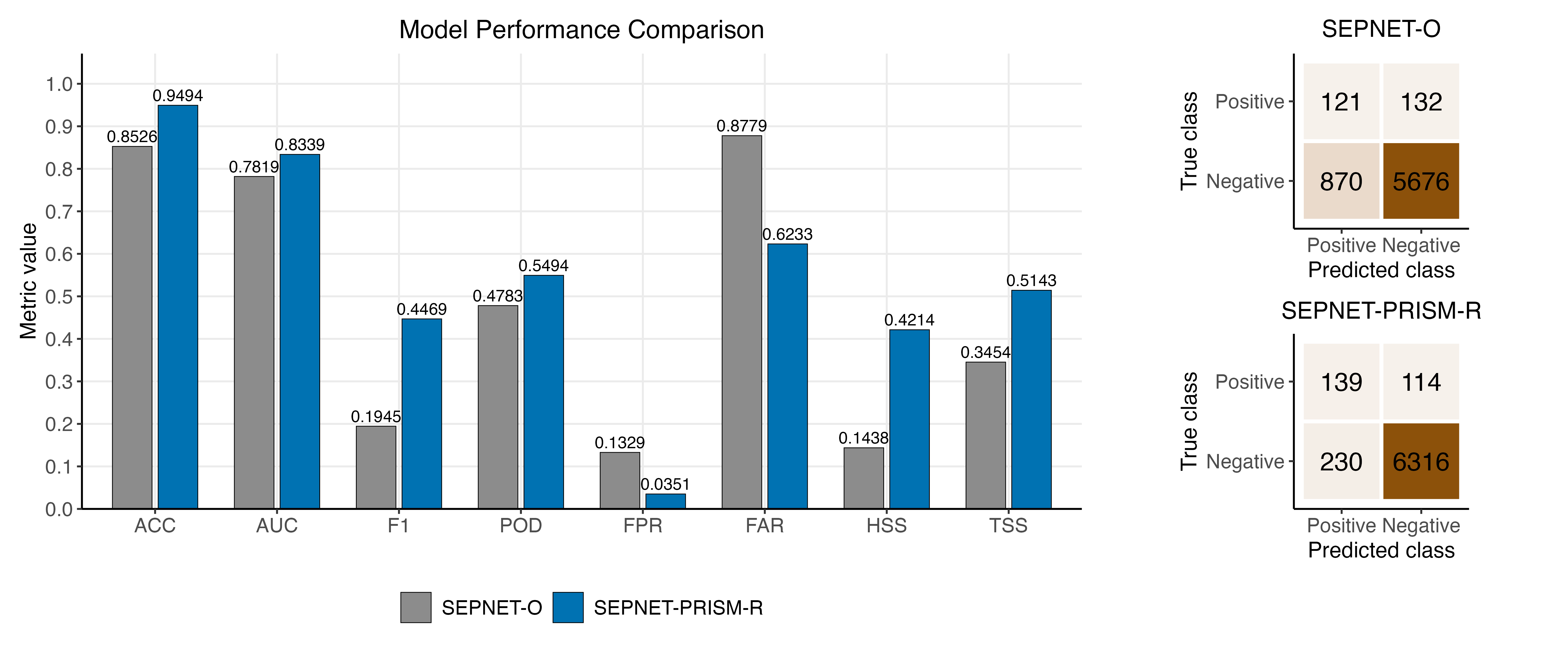}
        \caption{Summary of evaluation metrics and confusion matrices for \texttt{SEPNET-O} and \texttt{SEPNET-PRISM-R} over the full test period (September 2025 through July 2026).}
        \label{fig:Metrics_confusion_matrix}
    \end{subfigure}
    \caption{Overall performance of the operational SEP forecasting models.}
    \label{fig:Overall_results}
\end{figure*}

We evaluated the models over the full test period from September 2025 to July 2026. Figure \ref{fig:ProbsPFXRSB} shows the temporal evolution of forecast probabilities and warning states from \texttt{SEPNET-PRISM-R} and \texttt{SEPNET-O}, together with the corresponding proton flux and soft X-ray XRSB observations. Over this period, \texttt{SEPNET-PRISM-R} generated substantially fewer false alarms while still identifying the major SEP intervals of interest, whereas \texttt{SEPNET-O} issued frequent and often persistent warnings outside confirmed SEP intervals. Within this test period, three proton events met the operational SEP criterion of proton flux exceeding 10 pfu in the $>$ 10 MeV channel, beginning at 11:25 UTC on November 10, 2025, 22:55 UTC on January 18, 2026, and 00:15 UTC on May 26, 2026. The first two events were intense, with peak proton fluxes on the order of 3$\times$10$^3$ pfu, while the May 2026 event reached a much smaller peak of only about 20 pfu. It should be noted that the proton flux shown here is the unpreprocessed observational GOES measurement, and has not been background corrected \citep{fetchsep}.

As shown in Figure \ref{fig:ProbsPFXRSB}, \texttt{SEPNET-PRISM-R} successfully captured the first two high-intensity events and issued far fewer false warning periods than \texttt{SEPNET-O}. The remaining false alarms were associated with exceptionally strong soft X-ray flare activity, illustrating that intense solar eruptions do not always produce SEP events at Earth and underscoring the difficulty of forecasting rare SEP occurrences from precursor information alone. The aggregate verification statistics are summarized in Figure \ref{fig:Metrics_confusion_matrix}. Relative to \texttt{SEPNET-O}, \texttt{SEPNET-PRISM-R} improved all major metrics, increasing probability of detection (POD) from 0.4783 to 0.5494, HSS from 0.1438 to 0.4214, and TSS from 0.3454 to 0.5143, while reducing FPR from 0.1329 to 0.0351 and FAR from 0.8779 to 0.6233. The confusion matrices further show that false alarms decreased from 870 for \texttt{SEPNET-O} to 230 for \texttt{SEPNET-PRISM-R}, while correctly identified positive samples increased from 121 to 139. Collectively, these results indicate enhanced event detection performance and markedly improved control of false alarms for \texttt{SEPNET-PRISM-R}.

\subsection{SEP and Non-SEP Event Case Studies}\label{subsec2-2}

\begin{figure*}[htbp]
    \centering
    \begin{subfigure}[t]{1\textwidth}
        \centering
        \includegraphics[width=\textwidth]{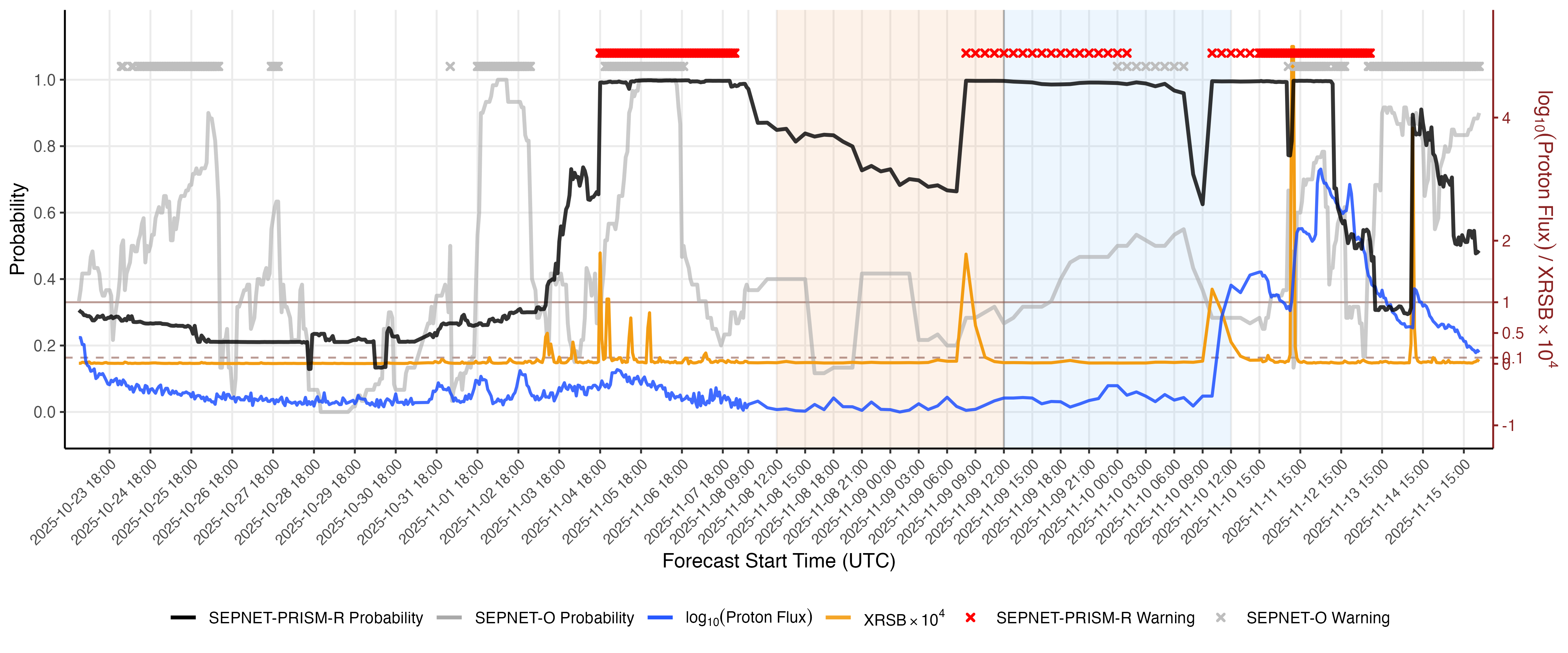}
        \caption{Forecast results for the November 2025 SEP event.}
        \label{fig:ProbsPFXRSB_Nov2025}
    \end{subfigure}
    \begin{subfigure}[t]{1\textwidth}
        \centering
        \includegraphics[width=\textwidth]{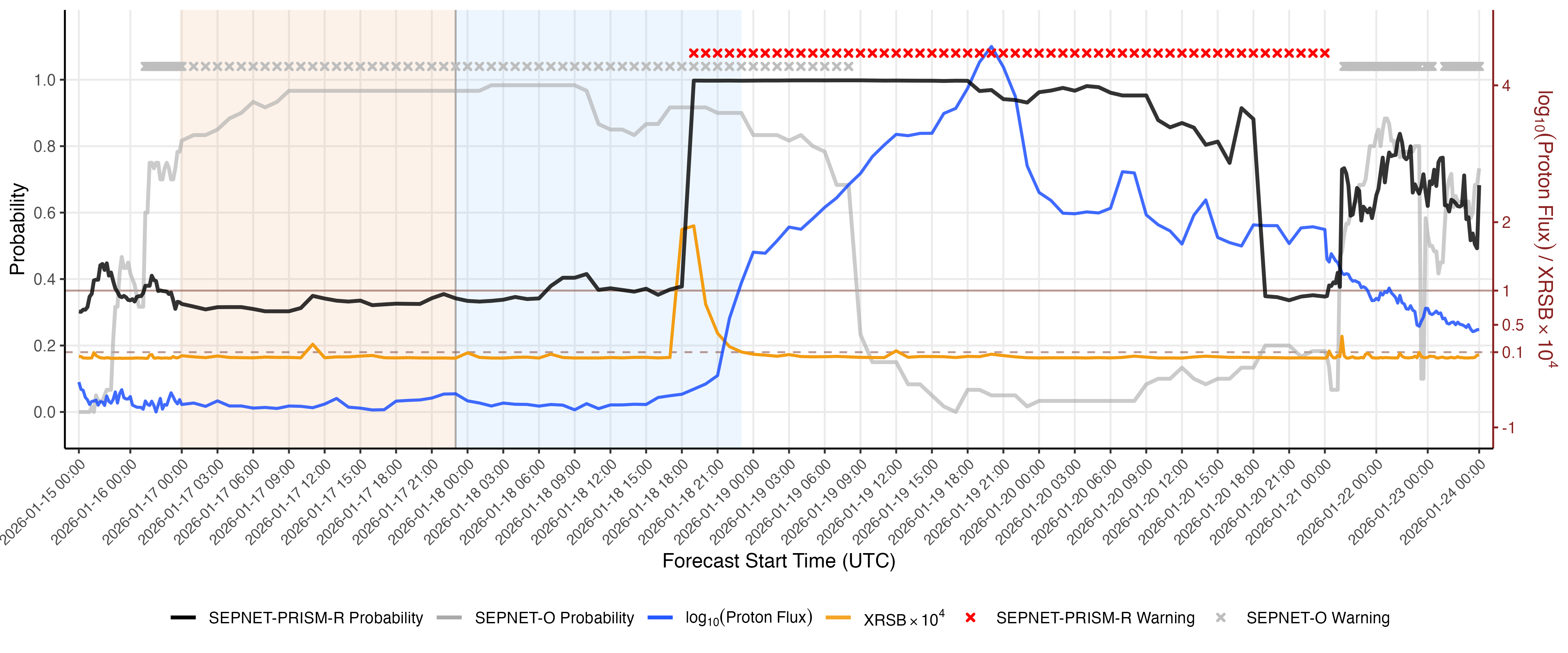}
        \caption{Forecast results for the January 2026 SEP event. }
        \label{fig:ProbsPFXRSB_Jan2026}
    \end{subfigure}
    \caption{Detailed visualizations of operational forecasts for two SEP events, November 2025 (top) and January 2026 (bottom). The left axis shows the forecast probabilities from \texttt{SEPNET-PRISM-R} (black) and \texttt{SEPNET-O} (gray), together with their corresponding warning indicators (red and gray crosses). The right axis shows $\log_{10}$ proton flux (blue) and soft X-ray flux XRSB scaled by $10^4$ (orange). The orange shaded interval denotes the 24-hour precursor window used to construct the forecast, and the blue shaded interval marks the forecast period preceding SEP onset.}
    \label{fig:Detailed_cases_SEP}
\end{figure*}

Figures \ref{fig:ProbsPFXRSB_Nov2025} and \ref{fig:ProbsPFXRSB_Jan2026} show detailed views of the two successfully forecasted SEP events. In both cases, \texttt{SEPNET-PRISM-R} issued warnings several hours before proton flux exceeded the 10 pfu threshold and maintained elevated probabilities throughout the main event interval. The orange shaded region denotes the 24-hour input window used to generate the forecast, and the blue shaded region denotes the forecast window preceding SEP onset. In the November 2025 event, the model responded to strong precursor activity, including a large soft X-ray enhancement, and issued a warning well before the proton flux crossed threshold. In the January 2026 event, the model again provided an early warning and sustained the alert throughout.

\begin{figure*}[htbp]
    \centering
    \begin{subfigure}[t]{1\textwidth}
        \centering
        \includegraphics[width=\textwidth]{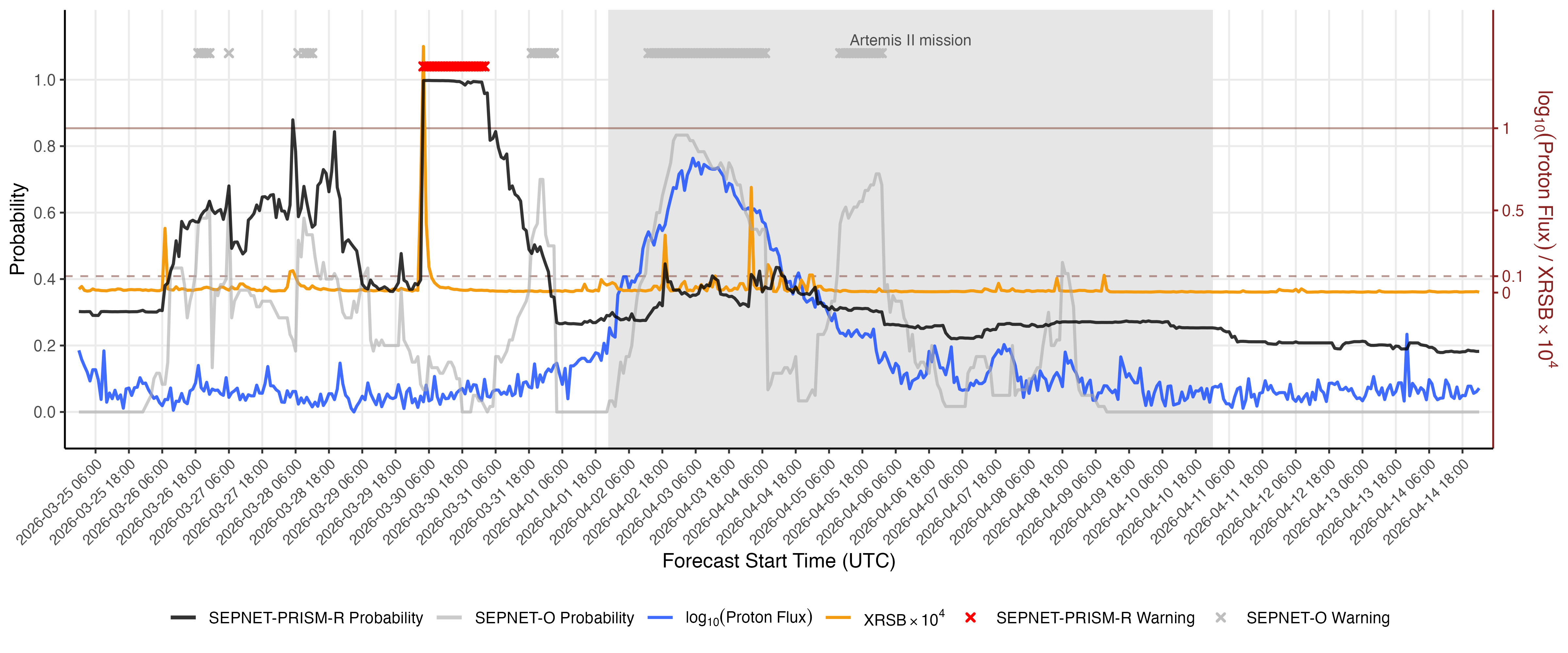}
        \caption{Forecast results during the Artemis II mission, including the extreme X-class flare that occurred on March 30, 2026.}
        \label{fig:ProbsPFXRSB_Apr2026}
    \end{subfigure}
    \begin{subfigure}[t]{1\textwidth}
        \centering
        \includegraphics[width=\textwidth]{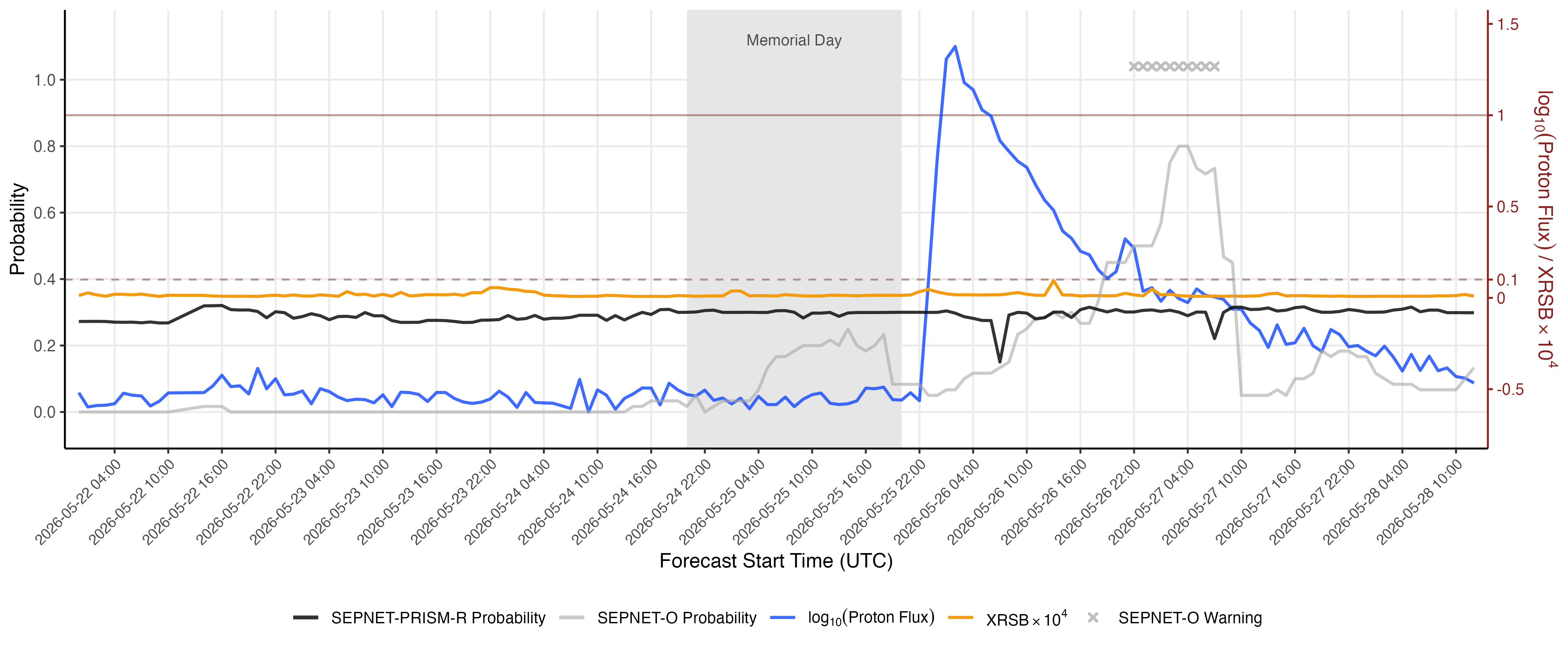}
        \caption{Forecast results during the late-May 2026 interval around Memorial Day.}
        \label{fig:ProbsPFXRSB_May2026}
    \end{subfigure}
    \caption{Detailed visualizations of operational forecasts during two disturbed intervals. The left axis shows the forecast probabilities from \texttt{SEPNET-PRISM-R} (black) and \texttt{SEPNET-O} (gray), together with their corresponding warning indicators. The right axis shows $\log_{10}$ proton flux (blue) and XRSB soft X-ray flux scaled by $10^4$ (orange). }
    \label{fig:Detailed_cases_NonSEP}
\end{figure*}

In addition to the verified SEP events, we examine two disturbed intervals to assess the model's robustness under challenging conditions. Figure \ref{fig:Detailed_cases_NonSEP} shows forecasting results during the Artemis II mission and the late-May 2026 event interval around Memorial Day. On March 30, 2026, an intense X-class flare occurred, and \texttt{SEPNET-PRISM-R} responded with elevated event probability (Figure \ref{fig:ProbsPFXRSB_Apr2026}) and gradually decayed over time. No SEP event was recorded during this interval. In contrast, \texttt{SEPNET-O} issued several scattered SEP warnings, indicating a strong false-alarm tendency under disturbed but non-SEP conditions. This case indicates that the updated framework, \texttt{SEPNET-PRISM-R}, is responsive to strong eruptive solar activity while suppressing spurious warnings when the disturbance does not evolve into a detectable SEP enhancement at Earth.  The late-May 2026 interval differs in that the proton enhancement was a true SEP event, likely driven by a CME, but it was not well represented by the short-horizon precursor variables used in both \texttt{SEPNET-O} and \texttt{SEPNET-PRISM-R}. Consequently, \texttt{SEPNET-PRISM-R} did not issue a timely warning, while \texttt{SEPNET-O} generated a delayed alert. This case suggests that CME-driven SEP events may require explicit CME-related inputs, motivating a more curated analysis of CME information in future work.

\subsection{Auxiliary Regression Results}\label{subsec2-3}
In addition to binary SEP event forecasting, \texttt{SEPNET-PRISM-R} predicts three auxiliary regression targets: the subsequent 24-hour maximums of logarithmic proton flux in the $>$ 10 MeV channel, X-ray flux XRSB, and the CME speed. These auxiliary tasks are intended to provide physically relevant information about the SEP environment, thereby enhancing feature learning and prediction accuracy. 

\begin{figure}[hpbt]
    \centering
\includegraphics[width=1\linewidth]{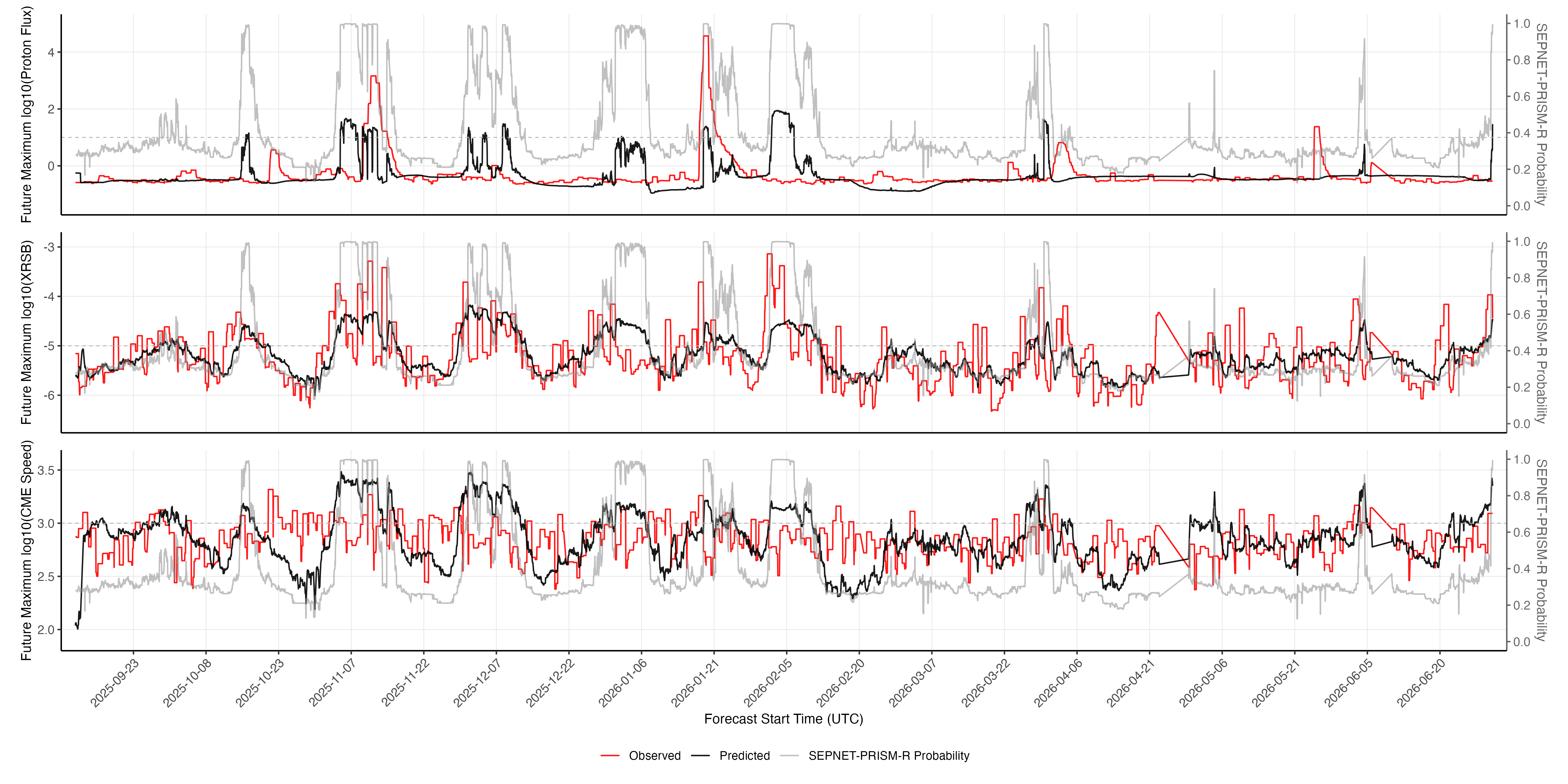}
\caption{Comparison between predicted and observed values for the two auxiliary regression targets. The upper panel presents the future maximum $\log_{10}$ proton flux in the $>$ 10 MeV channel, the middle panel presents the future maximum $\log_{10}$ XRSB flux, and the lower panel presents the future maximum $\log_{10}$ CME speed. The gray background curve indicates the estimated SEP-event probability from \texttt{SEPNET-PRISM-R} for reference.}
\label{fig:PFXRSBPred}
\end{figure}

Figure \ref{fig:PFXRSBPred} compares the predicted and observed values. The gray background curve indicates the estimated SEP event probability as a reference. In general, intervals of high estimated SEP probability are associated with elevated predicted proton flux, showing that \texttt{SEPNET-PRISM-R} has learned a physically meaningful relationship between SEP occurrence and enhanced proton intensity. At the same time, the predicted proton flux series remains somewhat elevated during some intervals in which the observed proton flux does not increase substantially, suggesting that accurate quantitative prediction of proton intensity remains challenging given the strongly skewed distribution of flux values. Moreover, the predicted X-ray flux XRSB follows the observed temporal variations more closely, whereas the skewed distributions resulting from the extremely high value are still not effectively captured. Accurately predicting the maximum CME speed remains challenging within the current framework, where CME speed is not the primary objective. This is also due to CME kinematics being influenced by complex, highly variable eruption dynamics, projection effects, and incomplete observability of input features \citep{Vourlidas2006,Zhao2024SOFIE}.

\subsection{SHAP Importance}\label{subsec2-4}
To interpret the operational model beyond the aggregate skill scores, we quantified the contribution of individual predictors to the probability of the calibrated SEP event using SHAP (SHapley Additive exPlanations) analysis \citep{wang2024feature,Daniel2022GBM}. For the preferred SHARP+flare (\texttt{S+F},\citep{yu2026SEPNET-PRISM}) predictor configuration of \texttt{SEPNET-PRISM-R}, we computed gradient explainer attributions for each forecast window across the ensemble of full-train split models, which were subsequently averaged. For clarity of presentation, we pooled the windowed summaries (minimum, maximum, and average) for each physical quantity into a single base feature. Additionally, we aggregated individual predictors into block totals for all SHARP and flare features.

\begin{figure*}[htbp]
    \centering
    \begin{subfigure}[t]{\textwidth}
        \centering
        \includegraphics[width=\textwidth]{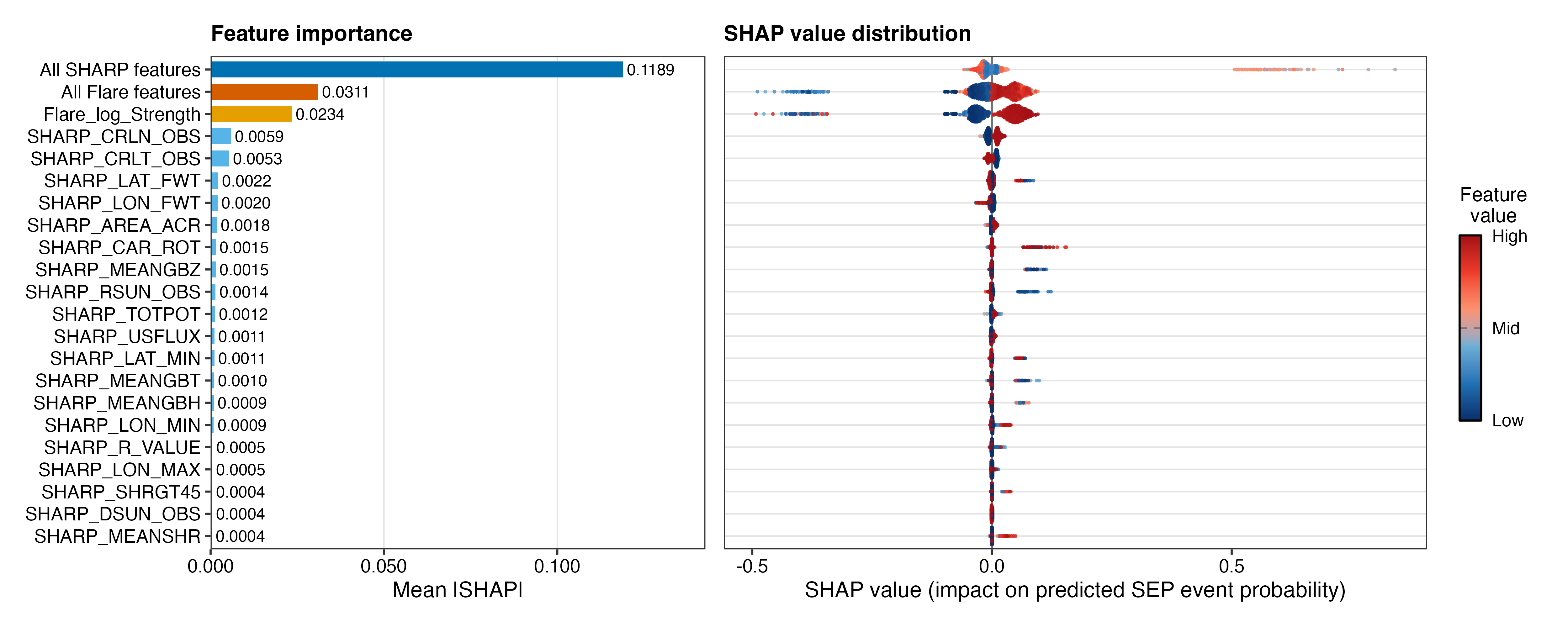}
        \caption{Global SHAP summary for \texttt{SEPNET-PRISM-R} with feature configuration \texttt{S+F}. Left: mean absolute SHAP importance for All-SHARP and All-Flare blocks and for the leading base features (pooling \texttt{min}/\texttt{max}/\texttt{avg}). Right: signed SHAP beeswarm colored by the corresponding feature value.}
        \label{fig:SHAP_summary}
    \end{subfigure}

    \vspace{0.6em}

    \begin{subfigure}[t]{\textwidth}
        \centering
        \includegraphics[width=\textwidth]{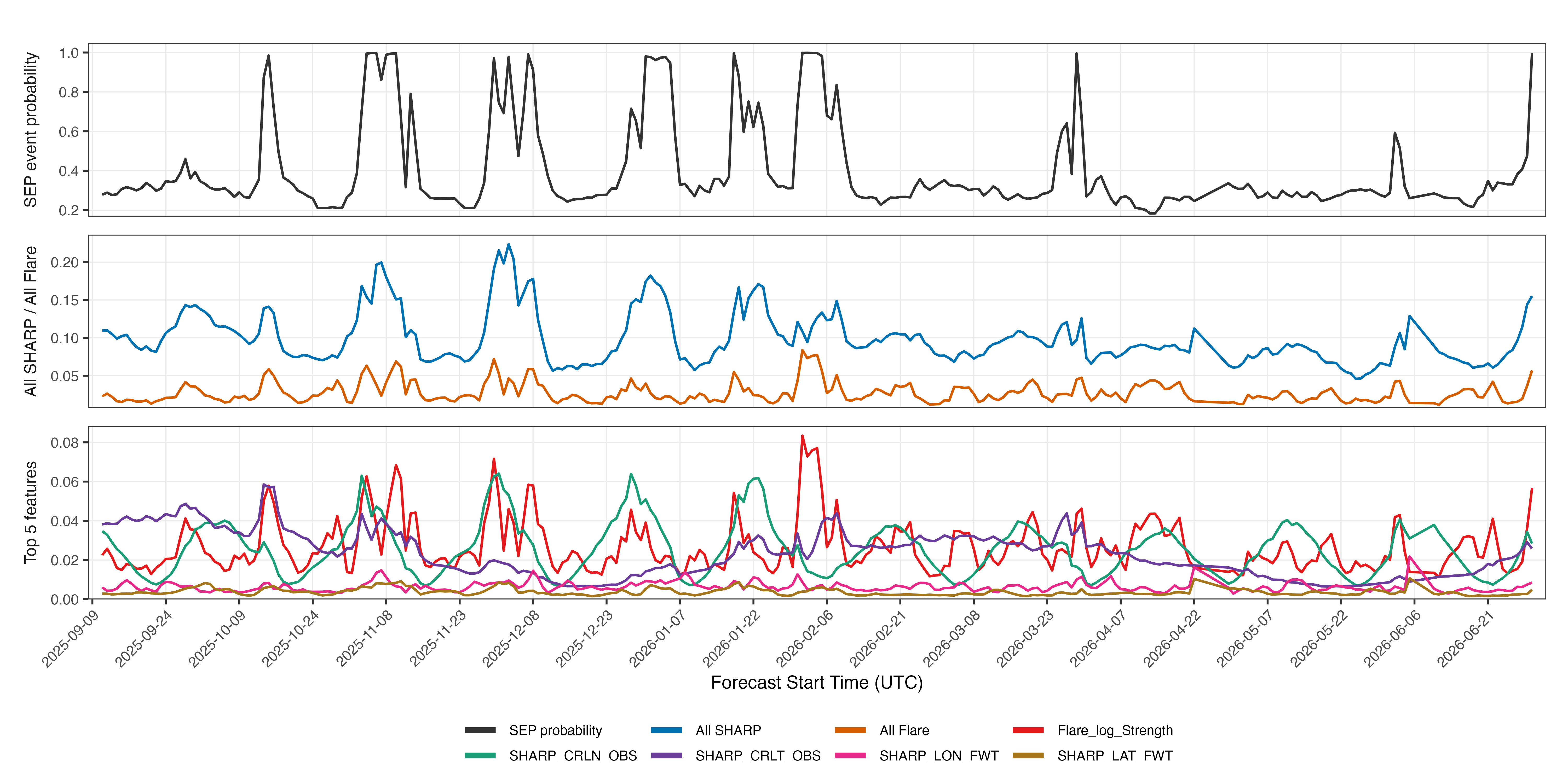}
        \caption{Time series of the ensemble SEP predicted probability (top), aggregate All-SHARP and All-Flare SHAP masses (middle), and leading individual base-feature contributions (bottom) over the near-real-time evaluation period. Values are daily means of $\Sigma|\mathrm{SHAP}|$ over \texttt{min}/\texttt{max}/\texttt{avg}; a few pathological post-outage days are omitted from the line plot.}
        \label{fig:SHAP_over_time}
    \end{subfigure}
    \caption{SHAP interpretation of \texttt{SEPNET-PRISM-R} under the operational \texttt{S+F} feature set. a) Global feature importance and signed contribution distributions. b) Temporal co-evolution of forecast probability and SHAP attributions.}
    \label{fig:SHAP}
\end{figure*}

The global feature importance summary is shown in Figure \ref{fig:SHAP_summary}. When importance is measured by the mean absolute attribution mass, the SHARP block dominates the flare block (mean $\Sigma|\mathrm{SHAP}|\approx 0.1189$ versus $\approx 0.0311$). Among individual base features, flare intensity (\texttt{Flare\_log\_Strength}) is the strongest single contributor (mean $|\mathrm{SHAP}|\approx0.0234$), followed by Carrington longitude and latitude of the observed region (\texttt{SHARP\_CRLN\_OBS}, \texttt{SHARP\_CRLT\_OBS}). Temporal consistency of these attributions is examined in Figure \ref{fig:SHAP_over_time}. In \ref{fig:SHAP_over_time}, the upper panel shows the ensemble SEP probability over the near-real-time test period; the middle panel compares the All-SHARP and All-Flare attribution masses; and the lower panel tracks the leading individual base features. Elevated SEP probabilities generally coincide with increases in the aggregate SHARP contribution and, at times, with flare-related attributions, supporting the interpretation that magnetic-context and flare information jointly drive the operational forecast of SEP events.

\subsection{A Web-Based Live Dashboard}\label{sec:dashboard}

The forecasting framework \texttt{SEPNET-PRISM-R} is deployed as a publicly accessible, hourly updated web dashboard, \texttt{SEPNET-PRISM-R} (\url{https://mlsw.engin.umich.edu/apps/SEPNET-PRISM-R}), hosted within the University of Michigan Space Weather Interactive Lab (SWIL, \url{spaceweatherlab.org}) prediction-model portal. The dashboard exposes six linked views, ``\textit{Probability}, \textit{Regression}, \textit{SHAP over time}, \textit{SHAP summary}, \textit{Similar period}, and \textit{Similar event}'', that together provide continuous access to the model outputs discussed in the preceding subsections, together with the analog-based contextualization introduced in Section~\ref{sec:analog}. The dashboard header reports the number of hourly forecasts, the number of active warnings, the covered forecast window, and the most recent update time. The underlying forecast archive is refreshed hourly and mirrored to a dedicated cloud storage location for reproducibility. 

The notable features of the web dashboard are its high interpretability and support for real-time decision-making. Typical real-time space weather forecasting websites issue warnings based on predicted event probability and pre-trained or preset decision thresholds. In our dashboard, we offer researchers a feature contribution decomposition to materialize the forecasting outcome in terms of physical quantities. Moreover, the raw model output, probability of a forthcoming event, does not directly give the reliability and trustworthiness of the forecast, and a preset threshold makes the decision-making arbitrary and sensitive to small perturbations of the system and noise contamination. We match the current solar condition with historical windows and historical events in the web interface, showing (i) how good/bad our model performed in similar solar cycle years and (ii) whether our model correctly/incorrectly predicted historical events that enjoyed similar features as the current event. For each one, we provide summary statistics so operators can easily rely on them when making operational decisions. 

\begin{figure*}[htbp]
    \centering
\includegraphics[width=1\linewidth]{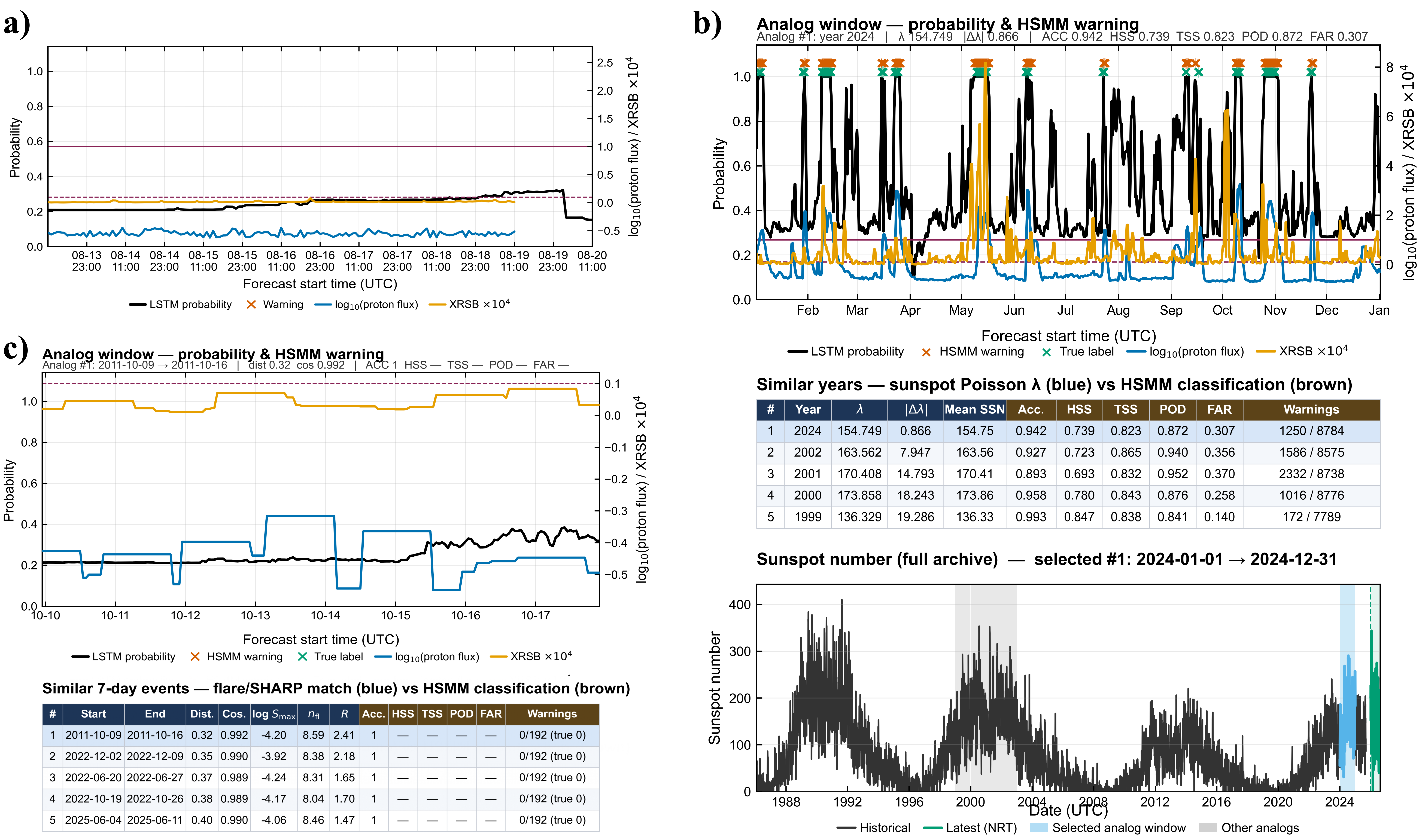}
    \caption{Representative views of the \texttt{SEPNET-PRISM-R} web dashboard. a) Real-time probability, HSMM warning state, and observed proton and soft X-ray flux. b) Similar solar-cycle periods, ranked by $|\Delta\lambda|$ in year-to-date sunspot-number Poisson rate, with historical HSMM classification skill for each analog year. c) Similar precursor events, ranked by z-scored Euclidean distance and cosine similarity in flare-and-SHARP feature space over non-overlapping 7-day windows.}
    \label{fig:dashboard}
\end{figure*}

Figure \ref{fig:dashboard} illustrates representative views from a live snapshot of the dashboard that gets continuously updated as new observations arrive each hour. The \textit{Probability view} (Figure \ref{fig:dashboard}a) gives the real-time forecast probability and the HSMM warning state alongside observed proton flux and soft X-ray flux, in the same style as Figure \ref{fig:ProbsPFXRSB}. The \textit{SHAP summary view} shows the global feature-importance ranking and beeswarm distribution described in Section \ref{sec2} in an interactive form, alongside a complementary \textit{SHAP-over-time view} that tracks the same attributions hour by hour. The \textit{Similar period view} (Figure \ref{fig:dashboard}b) ranks historical years by $|\Delta\lambda|$ relative to the current year's sunspot-number Poisson rate; in a representative snapshot from mid-August 2026 (Figure \ref{fig:dashboard}c), 2024 ($\lambda=154.7$, $|\Delta\lambda|=0.866$), 2002, 2001, 2000, and 1999 were returned as the top five analog years, with historical HSMM warning skill ranging from HSS $=0.72$ to $0.85$ across these analogs. The \textit{Similar event view} ranks historical non-overlapping 7-day windows by z-scored Euclidean distance in flare-and-SHARP feature space; for the same representative snapshot, the closest matched windows (e.g., 9--16 October 2011, 2--9 December 2022, 20--27 June 2022, 19--26 October 2022, and 4--11 June 2025) all had distances below 0.4 and cosine similarities around 0.99, and none coincided with an operational SEP event, consistent with the quiescent forecast probability observed during that period. Together, these analog views comparing historical events allow forecasters to cross-check the current model probability and its dominant SHAP drivers against both the broader solar-cycle context and specific historical analogs of the current precursor signature, supporting more informed operational decision-making than a point-probability forecast alone.

\section{Discussions on Model Performance}\label{sec3}

The primary contribution of this study is the transition of the \texttt{SEPNET} framework from retrospective model development to a realistic operational forecasting system. Unlike previous studies~\citep{yu2026SEPNET-PRISM}, \texttt{SEPNET-PRISM-R} was evaluated in a strictly post-training setting, where model parameters remained fixed, and forecasts were generated sequentially as new observations became available. This evaluation more closely represents how an operational forecasting system would perform in practice. The results show that the proposed framework substantially improves forecast reliability by reducing false alarms while maintaining strong detection capability, suggesting that temporal consistency is as important as classification accuracy for operational SEP forecasting.

The improved performance primarily stems from incorporating temporal information via the HSMM. Hourly forecasts are naturally correlated because consecutive prediction windows overlap considerably. Treating each forecast independently can therefore produce isolated probability fluctuations and fragmented warnings, like in \texttt{SEPNET-O}~\citep{yu2026SEPNET}. By explicitly modeling state persistence, the HSMM suppresses short-lived fluctuations while preserving sustained periods of elevated probability that are characteristic of true SEP events. This post-processing strategy is independent of the underlying neural network architecture and could therefore be combined with future forecasting models as a general framework for improving operational stability.

Beyond the point-probability forecast, the operational dashboard described in Section~\ref{sec:dashboard} is intended to broaden the basis on which forecasters interpret a given SEP alert. SHAP attributions indicate which physical drivers, chiefly SHARP magnetic-field parameters and flare intensity, are shaping a specific forecast, while the similar-period and similar-event analogs situate the current forecast within the broader solar-cycle context and against concrete historical precedents with known outcomes. This case-based layer does not alter the underlying probability but is intended to help forecasters assess the plausibility and historical precedent of a given warning, in a manner analogous to how ensemble and analog techniques are used to contextualize probabilistic forecasts in terrestrial weather prediction \citep{vandendool1989analog,dellemonache2013probabilistic}. 

The case studies also highlight several directions for future improvement. Although \texttt{SEPNET-PRISM-R} successfully forecasted the two major SEP events during the evaluation period, it failed to provide an early warning for the relatively weak event in May 2026. This event appears to have been more strongly associated with CME evolution than with the magnetic and flare precursors emphasized in the current model. Our supplementary analysis indicates that the available CME variables provide limited incremental predictive information beyond SHARP and flare parameters, likely because their current representation does not adequately capture eruption-specific characteristics. Future work shall therefore investigate more event-centered CME descriptions, together with longer operational evaluation over additional years of Solar Cycle 25, to further assess model robustness and generalization.

In summary, our numerical results indicate that combining multitask representation learning with explicit temporal post-processing and SHAP-based interpretability yields a forecasting system whose reliability and transparency are well-suited to operational deployment. By leveraging physically meaningful magnetic-field and flare-related predictors alongside an operational web dashboard that provides real-time probabilities, feature attributions, and historical analogs, this system represents a critical advancement toward reliable, interpretable, and continuous machine-learning architectures for space weather forecasting. 

\section{Methods}\label{sec4}

This section describes the methodology for developing the hourly near-real-time SEP forecasting framework and the SHAP-based feature importance analysis.

\subsection{Forecasting Design and Sample Construction}\label{subsec4-1}
\begin{figure}[hpbt]
    \centering
    \includegraphics[width=1\linewidth]{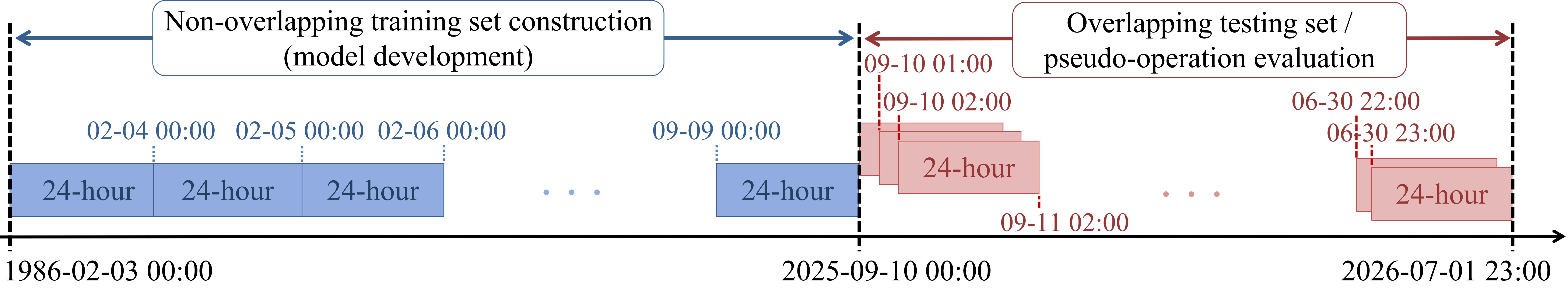}
    \caption{Illustration of the training and evaluation sample construction. Model development adopts non-overlapping 24-hour windows from February 1986 to September 2025 to avoid dependence and information leakage, while pseudo-operational evaluation uses hourly forecasts over an independent post-training period from September 10, 2025, to July 1, 2026.}
    \label{fig:DatasetConstruction}
\end{figure}

In the intended real-time forecasting setting, SEP predictions are issued hourly using predictor information collected over the preceding 24 hours. This design inherently produces input samples with substantial temporal overlap, such that consecutive samples share highly similar predictor histories. Constructing a full training dataset using one-hour-ahead sliding windows, therefore, induces strong dependence among samples. Under these conditions, a conventional random split into training and validation sets is not appropriate because neighboring samples are nearly identical, and can place highly similar instances in both sets, leading to information leakage. This leads to overly optimistic estimates of predictive performance and compromises the validity of generalization assessment. To address this issue, the training dataset for model development is constructed using non-overlapping 24-hour windows, as illustrated in Figure \ref{fig:DatasetConstruction}, following the procedure in \cite{yu2026SEPNET-PRISM} and \cite{yu2026SEPNET}, which enforces a clearer separation among samples and yields more reliable estimates of out-of-sample predictive skill.

The training data is constructed using observations from February 1986 through September 2025, consistent with the temporal coverage of the SEP benchmark dataset~\citep{whitman2026validation}. During this period, models are trained and validated across 50 random seeds to account for variability due to data partitioning and stochastic optimization. The resulting fitted models are then fixed and analyzed to evaluate the stability and reproducibility of predictive performance. To emulate near-real-time forecasting, a separate post-training interval beginning immediately after the end of the training period and continuing through 1 July 2026. This interval is excluded from all model training and tuning procedures and is treated as a pseudo-operational test period. Performance over this period is therefore intended to reflect how the forecasting system would operate if the model were to be adopted for real-time use.

The forecasting framework is built on the enhanced multitask deep learning model \texttt{SEPNET-PRISM}~\citep{yu2026SEPNET-PRISM}. At each forecast issue time, the model ingests solar observations from the preceding 24 hours, including active-region patch information, SHARP parameters, and solar-flare characteristics. In addition to predicting SEP occurrence, \texttt{SEPNET-PRISM} is designed as a multitask learning model that jointly forecasts three auxiliary quantities: the maximum logarithmic proton flux in the $>$ 10 MeV channel, the GOES soft X-ray flux, and the CME speed within the next 24 hours. These auxiliary targets are selected to capture key physical processes associated with SEP generation and to support shared representation learning across related tasks.

\subsection{Hidden Semi-Markov Model (HSMM) for Post-Processing}\label{subsec4-2}

At each forecast time, the previous model \texttt{SEPNET-PRISM} estimates the probability of an SEP event within the next 24 hours using observations from the previous 24-hour window. In near-real-time operation, forecasts are updated hourly using a sliding window, resulting in overlapping inputs and temporal dependence between successive predictions. While the neural network generates each probability independently, this overlap induces autocorrelation, leading to short-term fluctuations that are inconsistent with the typically continuous evolution of the process during SEP events. Consequently, probabilities near decision thresholds may produce fragmented or unstable event identification.

To mitigate this issue, \texttt{SEPNET-PRISM-R} incorporates a post-processing step based on an HSMM \citep{yu2010hidden}, which explicitly models temporal dependence via Markov transitions between states and state persistence via a sojourn-time distribution. Let $S_t$ denote the latent state at time $t$, where $S_t=0$ represents a quiet state and $S_t=d$, for $d=1,\ldots,D$, represents an SEP event with elapsed duration $d$ hours. Unlike standard hidden Markov models, the HSMM accounts for duration in each state, enabling a more realistic representation of event continuity. Integrating information over time produces a smoothed, temporally coherent sequence of event probabilities. To prevent information leakage, the HSMM is trained using only the original development period. An auxiliary dataset is constructed using the same hourly sliding-window procedure, and the trained and frozen \texttt{SEPNET-PRISM} model is then applied to generate predicted SEP probabilities. These serve as observations for HSMM training, allowing the model to learn the temporal dependence structure and uncertainty characteristics of the neural network outputs. Importantly, no data from the post-training evaluation period are used in this process, ensuring a fair assessment of the out-of-sample performance.

The HSMM takes as input the \texttt{SEPNET-PRISM} output of SEP event probability and physically motivated covariates derived from the preceding one hour before the forecast issue time, including the logarithmic maximum proton flux and GOES soft X-ray flux. Specifically, let $x_{\mathrm{prob},t}$ denote the standardized logit-transformed \texttt{SEPNET-PRISM} probability at time $t$, where the logit transform maps a probability $p_t\in (0,1)$ to the real line via $\log\{p_t / (1-p_t)\}$. Let $x_{\mathrm{pf},t}$ and $x_{\mathrm{cme},t}$ denote the logarithmic maximum proton flux and XRSB flux over the preceding one hour, respectively. These covariates are then used to estimate two time-varying hazards: an entry hazard that governs transitions from quiet to event, and an exit hazard that governs transitions from event back to quiet. The estimated entry hazard is given by
\[
\operatorname{logit}\{P(S_t>0 \mid S_{t-1}=0, \mathbf{x}_t)\}
=-8.06+1.43\,x_{\mathrm{prob},t}-0.74\,x_{\mathrm{pf},t}-0.07\,x_{\mathrm{xrs},t},
\]
which indicates that higher \texttt{SEPNET-PRISM} probabilities $x_{\mathrm{prob},t}$ increase the likelihood of transitioning from a quiet state to an event state, whereas the proton flux $x_{\mathrm{pf},t}$ and XRSB $x_{\mathrm{xrs},t}$ covariates exert weaker negative effects. The large negative intercept implies a low baseline entry probability unless the covariate evidence is sufficiently strong. The estimated exit hazard is 
\[
\operatorname{logit}\{P(S_t=0 \mid S_{t-1}=d, \mathbf{x}_t)\}
=0.54\,d_t-2.58
-0.40\,x_{\mathrm{prob},t}
-0.15\,x_{\mathrm{pf},t}
-0.31\,x_{\mathrm{xrs},t},
\]
where $d_t$ is the elapsed duration of the current event state. The positive duration coefficient indicates that longer event runs are increasingly likely to terminate, whereas stronger evidence from event-related covariates decreases the exit probability. This duration-dependent structure helps prevent unrealistically long event sequences and promotes more physically plausible event segmentation.

A weight parameter $w \in [0,1]$ is introduced to control the influence of the neural network output in the emission model (we selected $w=0.44$). Larger values of $w$ increase sensitivity but may raise false positives, while smaller values favor conservative predictions. The parameter is selected from the training set to maximize true positives while constraining false positives relative to the baseline model. In implementation, the HSMM expands event states into duration-specific substates up to a fixed maximum duration (72 hours). Forward--backward inference is used to compute smoothed posterior probabilities at each time step, and final event predictions are obtained by thresholding these probabilities.

\subsection{SHAP-Based Feature Importance Analysis}\label{subsec4-3}

To interpret the trained forecasting model, we conduct a SHAP-based feature importance analysis \citep{Daniel2022GBM,wang2024feature}. SHAP values quantify the marginal contribution of each predictor to the model output within a game-theoretic framework, enabling evaluation of individual features in a consistent, additive manner. We use SHAP analysis at both the global and local levels. At the global level, aggregated absolute SHAP values are used to rank predictors according to their overall contribution to the SEP forecasting model. At the local level, SHAP values are examined for individual forecasts to determine how specific input features increase or decrease the predicted probability of SEP. This analysis provides insight into the relative importance of active-region properties, magnetic-field parameters, and flare-related variables, thereby improving the interpretability and operational transparency of the proposed framework.

\subsection{Data Preparation}\label{sec:Data_Prep}

The primary prediction target in this study is the occurrence of operational SEP events, defined as intervals during which the proton flux exceeds 10 pfu in the $>$ 10 MeV energy channel. Event labels and events are obtained from the CLEAR SEP benchmark dataset through the \texttt{FetchSEP} python module \citep{fetchsep}, which provides event information spanning 3 February 1986 to 10 September 2025. A comprehensive description of data sources, preprocessing procedures, and quality control is provided in the dataset descriptor \citep{yu2026SEP-PRISMData}, with the dataset publicly archived on Zenodo \citep{yu2026SEP-PRISM_Data_zenodo}. Following prior work \citep{yu2026SEPNET-PRISM}, several groups of candidate predictors were considered, including magnetic-field parameters, flare-related variables, CME-related variables, soft X-ray flux, and historical proton flux. Comparative analyses indicate that magnetic-field parameters derived from SHARP-consistent data products, together with flare-related variables, provide the most stable predictive performance. In contrast, CME-related variables contribute less consistently, as detailed in the Supplementary Information. Although CME features are physically relevant to SEP production, our analysis shows that they are substantially redundant with the SHARP-based magnetic and flare predictors and provide only limited incremental value for the operational classification task. Historical proton flux can improve short-term event detection but can encourage the model to rely on immediate precursor signatures, thereby reducing the effective forecast lead time. Accordingly, the final predictor set is restricted to magnetic-field and flare-related features, aligning with the \texttt{SEPNET-O} model in \cite{yu2026SEPNET}.

All predictors are drawn from the curated, multi-source dataset and are represented in a temporally aligned and quality-controlled format. To construct supervised learning samples, predictor variables are aggregated over the 24-hour period preceding each forecast issuance time. For each variable, summary statistics (minimum, mean, and maximum) are computed to characterize recent solar conditions. These predictors are paired with targets defined over the subsequent 24-hour forecast horizon, including a binary indicator of SEP occurrence and auxiliary regression targets for proton flux, soft X-ray flux, and CME speed. This formulation yields a consistent dataset for joint classification and regression modeling of SEP-related activity.

\subsection{Analog-Based Contextualization: Similar Solar-Cycle Periods and Similar Precursor Windows}\label{sec:analog}

To complement the forecast and support situational interpretation, the deployment of \texttt{SEPNET-PRISM-R} also provides two forms of analog-based context: (i) historical years with a comparable level of solar-cycle activity to the ongoing year, and (ii) historical 7-day precursor windows whose flare and magnetic-field characteristics closely resemble the current forecast window. Both analogs are automatically recomputed at each hourly dashboard refresh and are intended to provide forecasters with an empirical, case-based reference for interpreting current model output, complementing the model-based probability and the SHAP attributions described above.

The first analog situates the current forecast within the broader solar cycle. For the current year-to-date, a Poisson rate $\lambda$ is fit to the daily sunspot number, providing a compact summary of the ambient level of solar activity, and each historical year in the training archive (1986 to present) is assigned an analogous $\lambda$ estimated over the same fitting window. Candidate years are ranked in ascending order of $|\Delta\lambda|$, the absolute difference between the current year's $\lambda$ and each historical year's $\lambda$, and the five years with the smallest $|\Delta\lambda|$ are reported as similar periods. For each analog year, the realized classification performance of the HSMM-based warning system, evaluated against the operational SEP label for that year (ACC, HSS, TSS, POD, and FAR), is also reported, allowing users to assess how the forecasting system historically performed during a comparably active/inactive solar-cycle period.

A second, independent analog examines the shape of the immediate-precursor input features. The dashboard identifies historical, non-overlapping 7-day windows whose flare-related features and SHARP parameters summaries most closely resemble those of the current 7-day forecast window. Candidate windows are drawn from the full historical archive and ranked using a z-scored Euclidean distance computed over flare-activity summaries (e.g., flare count and maximum and mean logarithmic peak flux) and SHARP magnetic-field summaries (e.g., total unsigned flux, the $R$-value, and mean shear angle); cosine similarity between the same feature vectors is reported alongside the distance as a complementary, scale-invariant measure of directional agreement. The five nearest historical windows are displayed as similar events, along with the realized SEP outcome and HSMM-based classification performance for each matched window, providing an event-level, case-based complement to the aggregate SHAP-based interpretation and allowing users to examine specific historical analogs of the current precursor configuration.

\backmatter

\bmhead{Supplementary information}

Supplementary Information contains additional details on the CME-related predictor comparison referenced in Section \ref{sec:Data_Prep}.

\bmhead{Acknowledgements}

The authors thank Ian Richardson (University of Maryland, NASA GSFC) and A. Steve Johnson (Leidos, NASA JSC SRAG) for their efforts in acquiring and preparing the dataset used in this study. 

\section*{Declarations}

\bmhead{Funding}
This work is supported by the NASA Space Weather Center of Excellence program under award No.\ 80NSSC23M0191 and No.\ 80NSSC23M0192.

\bmhead{Conflict of interest/Competing interests}
The authors declare no competing interests.

\bmhead{Ethics approval and consent to participate}
Not applicable. This study did not involve human participants, human data, or animal subjects.

\bmhead{Consent for publication}
Not applicable.

\bmhead{Data availability}
The training dataset underlying \texttt{SEPNET-PRISM-R} is archived on Zenodo \citep{yu2026SEP-PRISM_Data_zenodo}, and the SEP event benchmark used to construct it is described in \cite{fetchsep} and accessible through the \texttt{FetchSEP} python module. The operational forecasts generated by \texttt{SEPNET-PRISM-R}, including real-time event probabilities, auxiliary regression outputs, SHAP-based feature attributions, and the similar-period and similar-event analogs described in Methods, are made publicly available through an hourly updated web dashboard at \url{https://mlsw.engin.umich.edu/apps/SEPNET-PRISM-R}, hosted by the University of Michigan Space Weather Interactive Lab (SWIL).

\bmhead{Materials availability}
Not applicable. This study did not generate new physical materials.

\bmhead{Code availability}
The code implementing the \texttt{SEPNET-PRISM-R} forecasting framework, including model training, the HSMM temporal post-processing layer, and the SHAP-based interpretability analysis, is publicly available at \url{https://github.com/yuyian/SEPNET-PRISM-R}.

\bmhead{Author contribution}
Y.Y. and Y.C. conceived the study. Y.Y. conducted the data preparation, model development, and experiments. Y.Y. and Y.C. analyzed the results and wrote the manuscript. L.Z., K.W., W.M., and T.G. contributed to the scientific interpretation and revision of the manuscript. All authors reviewed the manuscript.

\bibliography{sn-bibliography}

\end{document}